\documentstyle[aps,prl,twocolumn,epsf]{revtex}

\newcommand{\beq}{\begin{equation}}
\newcommand{\eeq}{\end{equation}}
\newcommand{\bea}{\begin{eqnarray}}
\newcommand{\eea}{\end{eqnarray}}
\begin{document}

\title{ Quantum corrections to the ground state energy of
inhomogeneous neutron matter }

\author{ Aurel BULGAC $^{1,2}$ and Piotr MAGIERSKI $^{1,3}$}

\address{ $^1$Department of Physics,  University of Washington,
Seattle, WA 98195--1560, USA  }
\address{$^2$ Max--Planck--Institut f\"ur Kernphysik, Postfach 10 39
  80, 69029 Heidelberg, GERMANY}
\address{$^3$ Institute of Physics, Warsaw University of Technology,
  ul. Koszykowa 75, PL--00662, Warsaw, POLAND }

\date{\today }

\maketitle
\begin{abstract}

We estimate the quantum corrections to the ground state
energy in neutron matter (which could be termed as well either
shell correction energy or Casimir energy) at
subnuclear densities, where various types of inhomogeneities (bubbles,
rods, plates) are energetically favorable. We show that the magnitude
of these energy corrections are comparable to the energy differences
between various types of inhomogeneous phases. We discuss the
dependence of these corrections on a number of physical parameters
(density, filling factor, temperature, lattice distortions).

\end{abstract}

% 21.10.Dr -- Binding energies and masses,
% 21.65.+f -- Nuclear matter
% 97.60.Jd -- Neutron stars

%----------------------------------------------------------------------
\vspace{0.5cm}
\narrowtext

The investigation of the nuclear matter in the neutron star crust
below the saturation density leads to the consideration of exotic
shapes of the nuclei immersed in a neutron gas.  It was realized long
ago \cite{baym} that when the nuclei in dense matter occupy more than
half of the space it is energetically favorable to ``turn the nuclei
inside out'' and form a bubble phase \cite{rav,hash,oya}. To date a
large number of calculations have been performed pertaining to the
structure of the neutron star crust. The liquid drop model or the
Thomas--Fermi approximation calculations
\cite{lat,wil,las,oya1,lor,pet,wat} predict rather small energy
differences between different phases, of the order of a few
$keV/fm^{3}.$ (N.B. even though we often refer to energy, we actually
mean energy density.).  Apparently an agreement has been reached
concerning the existence of the following chain of phase changes as
the density is increasing: nuclei $\rightarrow$ rods $\rightarrow$
plates $\rightarrow$ tubes $\rightarrow$ bubbles $\rightarrow$ uniform
matter.  The density range for these phase transitions is $0.04 - 0.1
fm^{-3}$ \cite{lor,pet}. Moreover, it was established that these
phases exist up to temperatures of about $10 MeV$ \cite{lat}. At
densities of the order of several nuclear densities the quark degrees
of freedom get unlocked and the formation of various quark matter
droplets embedded in nuclear matter becomes then energetically
favorable \cite{heiselberg}.

The appearance of different phases is attributed to the interplay
between the Coulomb and surface energies.  Since most of the published
works were based on the minimization of some density functional in a
single Wigner--Seitz cell, the calculation of the shell correction or
Casimir energy has been omitted. In Hartree--Fock calculations
\cite{negele} these quantum corrections to the ground state energy of
neutron matter are obviously automatically incorporated. The
Hartree--Fock calculations performed so far were limited to
``spherical Wigner--Seitz cells'', which is arguably a reasonable
approximation for the ``nuclei in neutron gas'' phase only.  To our
knowledge there exist only one study on this subject where the shell
effects due to the bound nucleons only however (mainly protons) have
been taken into account \cite{oya2}.  It was determined that the shell
correction energy is smaller than the energy difference between
different phases and it was thus concluded that quantum corrections to
the ground state energy will not lead to any qualitative changes in
the sequence of the nuclear shape transitions in the neutron star
crust.

Our goal is to reach a comprehensive understanding of the so called
shell correction or Casimir energy in neutron stars. There is no well
established terminology for the energy corrections we are considering
here, even though the problem has been addressed before to some extent
by other authors. In the case of finite systems, the energy difference
between the true binding energy and the liquid drop energy of a given
system is typically refered to as shell correction energy. In field
theory a somewhat similar energy appears, due to various fluctuation
induced effects and it is generically referred to as the Casimir
energy \cite{casimir}:
%----------------------------------------------------------------------
\beq
E_{Casimir} = \int _{-\infty }^\infty d\varepsilon
\varepsilon [g(\varepsilon ,{\bf l} )-g_0(\varepsilon )],
\eeq
%----------------------------------------------------------------------
where $g_0 (\varepsilon )$ is the density of states per unit volume
for the fields in the absence of any objects, $g (\varepsilon ,{\bf
l})$ is the density of states per unit volume in the presence of some
``foreign''objects, such as plates, spheres, etc., and ${\bf l}$ is an
ensemble of geometrical parameters describing these objects and their
relative geometrical arrangement. A similar formula can be written for
neutron matter energy
%----------------------------------------------------------------------
\beq
E_{nm}=
 \int_{-\infty}^ \mu    d\varepsilon\varepsilon g  (\varepsilon ,{\bf l})
-\int_{-\infty}^{\mu _0}d\varepsilon\varepsilon g_0(\varepsilon ,{\bf l}),
\eeq
%----------------------------------------------------------------------
with the notable difference in the upper integration limit.  In the
above equation $g _0(\varepsilon , {\bf l})$ stands for the
Thomas--Fermi or liquid drop density of states of the inhomogeneous
phase and $g (\varepsilon ,{\bf l} )$ for the true quantum density of
states in the presence of inhomogeneities.  The parameters: $\mu$ and
$\mu _0$ are determined from the requirement that the system has a
given average density
%----------------------------------------------------------------------
\beq
\rho =\int _{-\infty }^\mu d\varepsilon  g (\varepsilon ,{\bf l} )
=  \int _{-\infty }^{\mu _0} d\varepsilon g _0(\varepsilon ,{\bf l} ).
\eeq
%----------------------------------------------------------------------
Since in infinite matter the presence of various inhomogeneities does
not lead to the formation of discrete levels, one might expect to
refer to corresponding energy correction for neutron matter as the
Casimir energy. In Ref. \cite{oya2} the authors computed a somewhat
different quantity however, than the one we are interested in this
work, the correction to the ground state energy arising from existence
of almost discrete levels inside a nucleus in an infinite
medium. Strictly speaking these levels are not discrete, but form
narrow energy bands due to the tunneling between neighboring
nuclei. The effects we shall consider here arise from the ``outside''
states, which is in complete analogy with the procedure for computing
the Casimir energy. As we shall show, these energy corrections,
arising from the existence of these truly delocalized states, are
larger than those considered in Ref. \cite{oya2}.  We have considered
similar issues earlier in finite systems and to some extent in
infinite 2--dimensional systems as well in Refs. \cite{bub1,bub2}.

In order to better appreciate the nature of the problem we are
addressing in this work, let us consider the following situation. Let
us imagine that two spherical identical bubbles have been formed in an
otherwise homogeneous neutron matter. For the sake of simplicity, we
shall assume that the bubbles are completely hollow.  We shall
sidestep the question of stability of each bubble for the moment and
assume that they are stable and rigid as well. We shall ignore the
role of long range forces, namely the Coulomb interaction in the case
of neutron stars, as their main contribution is to the smooth, liquid
drop or Thomas--Fermi part of the total energy.  Under such
circumstances one can ask the following apparently innocuous question:
``What determines the most energetically favorable arrangement of the
two bubbles?'' According to a liquid drop model approach (completely
neglecting for the moment the possible stabilizing role of the Coulomb
forces) the energy of the system should be insensitive to the relative
positioning of the two bubbles. A similar question was raised in
condensed matter studies, concerning the interaction between two
impurities in an electron gas. In the case of two ``weak'' and
point--like impurities the dependence of the energy of the system as a
function of the relative distance between the two impurities ${\bf a}$
is given by (spin coordinates are not displayed)
%--------------------------------------------------------------------
\begin{equation}
E({\bf a})=\frac{1}{2}\int d{\bf r}_{1} \int d{\bf r}_{2}
V_{1}({\bf r}_{1})\chi ({\bf r}_{1}-{\bf r}_{2}-{\bf a})V_{2}({\bf r}_{2}),
\end{equation}
%--------------------------------------------------------------------
where $\chi ({\bf r}_{1}-{\bf r}_{2}-{\bf a}) $ is the Lindhard
response function of a homogeneous Fermi gas and $V_{1}({\bf r}_{1})$
and $V_{2}({\bf r}_{2})$ are the potentials describing the interaction
between impurities and the surrounding electron gas.  At large
distances $k_{F}a \gg 1$ this expression leads to the interaction
first derived by Ruderman and Kittel \cite{rk,fw}:
%--------------------------------------------------------------------
\begin{equation} \label{r-k}
E ({\bf a}) \propto \frac{\hbar ^2}{2 m k_{F}a^{3}} \cos (2
k_{F} a),
\end{equation}
%--------------------------------------------------------------------
where $k_{F}$ is the Fermi wave vector and $m$ is the fermion mass
%--------------------------------------------------------------------
\begin{equation}
\mu=\frac{\hbar ^2k_F^2}{2 m}.
\end{equation}
%--------------------------------------------------------------------
This asymptotic behavior is valid only for point--like impurities,
when $k_{F}R \ll 1$, and where $R$ stands for the radius of the two
impurities. This condition is typically violated for either nuclei
embedded in a neutron gas or bubbles, when typically $k_FR\gg 1$. As
we shall show, in the case of large ``impurities'' (when $k_FR\gg 1$)
the interaction energy changes in a rather dramatic manner.  If one
replaces the ``weak'' impurities with ``strong'' point--like
impurities, only the magnitude of the interaction changes at large
distances, but not the form \cite{bub2}. The interaction (\ref{r-k})
has a pure quantum character, and any ``noise'' (e.g. temperature)
leads to a quick disappearance of the oscillatory behavior and with it
of the power law character, and the regular Debye screening (which is
exponential in character) sets in instead.

The lesson one can learn from this analysis however, is that quantum
corrections are most likely responsible for the interaction of two
bubbles/nuclei embedded in a Fermi gas and the form of the interaction
(\ref{r-k}) suggests the most natural way to proceed.  The argument of
the cosine is nothing else but the classical action in units of
$\hbar$ of the bouncing periodic orbit between the two impurities.
The exact form and magnitude of the coefficient in front of the cosine
can be obtained in a semiclassical approximation only after a careful
estimation of the leading order correction to the leading
semiclassical result.  Using the 3--dimensional extension of the
semiclassical approximation to the so called small disks problem
\cite{small}, we were able to obtain a significantly simpler and more
transparent derivation of this interaction than the original
derivation \cite{fw} as follows. The correction to the
single--particle propagator, which depends on the presence of the two
weak widely separated point--like impurities ($R/a\ll 1$ and $k_FR\ll
1$) is
%--------------------------------------------------------------------
\begin{eqnarray}
 \delta G  ({\bf r}  , {\bf r}^\prime, k) &\propto &
        G_0({\bf r}  , {\bf r}_1,      k)
        G_0({\bf r}_1, {\bf r}_2,      k)
        G_0({\bf r}_2, {\bf r}^\prime, k) \nonumber \\
 & + &  G_0({\bf r}  , {\bf r}_2,      k)
        G_0({\bf r}_2, {\bf r}_1,      k)
        G_0({\bf r}_1, {\bf r}^\prime, k),
\end{eqnarray}
%--------------------------------------------------------------------
where
%--------------------------------------------------------------------
\begin{equation}
G_0({\bf r}_1, {\bf r}_2, k)=-
\frac{m\exp ( ik|{\bf r}_1-{\bf r}_2|)}{2\pi\hbar ^2|{\bf r}_1-{\bf r}_2|}
\end{equation}
%--------------------------------------------------------------------
is the free single--particle propagator. Since only ``periodic
orbits'' contribute to the density of states, the correction to the
density of states, due to the presence of the two impurities and which
depends on their relative separation only is given by
%--------------------------------------------------------------------
\begin{equation}
\delta g (k,|{\bf r}_1-{\bf r}_2| ) \propto
{\mathrm{Im}}\left [
\frac{i\exp ( 2i|{\bf r}_1-{\bf r}_2|)}{k|{\bf r}_1-{\bf r}_2|}
\right ]
\end{equation}
%--------------------------------------------------------------------
and the corresponding correction to the ground state energy is given
by the obvious formula
%--------------------------------------------------------------------
\begin{eqnarray}
\delta {\cal{E}} (|{\bf r}_1-{\bf r}_2|)& \propto &
\int _{k\le k_F} kdk
                 \left ( \frac{\hbar ^2 k^2}{2m}-\mu \right )
\delta g (k,|{\bf r}_1-{\bf r}_2| ) \nonumber \\
& \propto &
\frac{\cos ( 2k_F|{\bf r}_1-{\bf r}_2|)}{|{\bf r}_1-{\bf r}_2|^3} .
\end{eqnarray}
%--------------------------------------------------------------------
Only the leading term in the limit $|{\bf r}_1-{\bf r}_2|\rightarrow
\infty$ is explicitly shown here. The proportionality coefficient is
naturally determined by the impurity strength.

The formation of various inhomogeneities in an otherwise uniform Fermi
gas can be characterized by several natural dimensionless parameters,
$k_Fa\gg 1$, where as above $a$ is a characteristic separation
distance between two such inhomogeneities, $k_FR\gg 1$, where $R$ is a
characteristic size of such an inhomogeneity, and $k_Fs \approx 1$,
where $s$ is a typical ``skin'' thickness of such objects. The fact
that the first two parameters, $k_Fa$ and $k_FR$, are both very large
makes a semiclassical approach natural. Since the third parameter,
$k_Fs$, is never too large or too small, one might be tempted to
discard a semiclassical treatment of the entire problem
altogether. However, there is a large body of evidence pointing
towards the fact that even though this parameter in real systems is of
order unity, the approximation $k_Fs\ll 1$, which we shall adopt in
this work, is surprisingly accurate \cite{brack}.  Moreover the
corrections arising from considering $k_Fs = {\cal O}(1)$ should lead
to an overall energy shift mainly, which is largely independent of the
separation among various objects embedded in a Fermi gas.  On one
hand, this type of shift can be accounted for in principle in a
correctly implemented liquid drop model or Thomas--Fermi
approximation. On the other hand, the semiclassical corrections to the
ground state energy arising from the relative arrangement of various
inhomogeneities have to be computed separately, as they have a
different physical nature. We are thus lead to the natural assumption
that a simple hard--wall potential model for various types of
inhomogeneities appearing in a neutron Fermi gas is a reasonable
starting point to estimate quantum corrections to the ground state
energy, see Refs. \cite{bub1,bub2,brack,caio} and earlier references
therein. We shall refer to these quantum corrections to the energy as
shell effects in the rest of the paper.  One might expect that such
simplifications will result in an overestimation of the magnitude of
shell effects, but the qualitative pattern should remain the same.

We shall consider spherical bubble--like, rod--like and plate--like
phases only here and we shall estimate the shell correction or Casimir
energy arising due to a regular arrangement of such inhomogeneities in
an otherwise homogeneous neutron gas. One can distinguish two types of
``bubbles'': {\it i)} nuclei--like structures embedded in a neutron
gas and {\it ii)} void--like structures. By voids we mean the regions
in which the nuclear density is significantly lower than in the
surrounding space.  In the first case {\it i)}, the single particle
wave functions can be separated into roughly two classes, those
localized mostly inside the nuclei--like structures and those which
are completely delocalized. A fermion in a delocalized state will
spend some time inside the ``nuclei'' too, but since the potential
experienced by a nucleon is deeper there, the local momentum is larger
and thus the relative time and relative probability to find a nucleon
in this region is smaller. One can approximately replace then the
``nuclei'' with an effective repulsive potential of roughly the same
shape. In the case of a ``bubble'', when the probability to find a
nucleon inside a ``bubble'' is reduced, again such an approximation
appears as reasonable. The ``nuclei'' and ``bubbles'' we are refering
to here, are not necessarily spherical, but could have the shape of a
rod or plate as well. There are of course a number of ``resonant''
delocalized states, whose amplitude behaves in a manner just opposite
to the one we have described here. However, the number of such
``resonant'' states is typically small and we thus do not expect large
effects due to them. Moreover, since such states are concentrated
mostly inside a ``nucleus'' or a ``bubble'' one does not expect them
to affect in a major way the relative positioning of two ``nuclei'' or
two ``bubbles''.  Nevertheless, these are some issues, which certainly
deserve more scrutiny in the future, even though we hardly expect that
a more comprehensive analysis will lead to qualitative changes of our
conclusions. In all these phases the shell effects depend on the
structure and stability of periodic orbits in the system
\cite{brack}. Except for the plate--like nuclei phase, where the shell
energy can be computed exactly, for other geometries one should
calculate the contribution from all periodic orbits. This is rather
tedious task, since they proliferate exponentially as a function of
their length \cite{pri}, and moreover this is not really necessary to
perform. If one is interested in the gross structure only of the shell
effects, the contribution of the shortest periodic orbits should
suffice for defining the gross shell structure. (We remind the reader,
in an infinite medium there are really no shells as in a finite
system, but we refer to the corresponding effects in this manner only,
due to their similar origin because of the appearance of periodic
orbits.)  Since the contribution of any given periodic orbit leads to
an oscillatory contribution to the density of states any suitably chosen
energy averaging over the spectrum, and in particular
a finite temperature as well, will leave only the contributions due to
the shortest periodic orbits. Since a periodic orbit of length
${\cal{L}}$ will lead to detail on an energy scale of the order
$\Delta E = \hbar^2 \pi^2/2m{\cal{L}}^2$, performing an averaging over
an energy interval $\Delta E $ will effectively mask the contribution
of orbits of length ${\cal{L}}$ or larger.  Moreover, since the
geometry of the rod--like and spherical phases admit only unstable
(hyperbolic) orbits, the longer the orbits, the lesser their
contribution is, due their decreased stability.

The simplest system consist of plate--like nuclei with the neutron gas
filling the space between slabs.  The shell energy for this system per
unit volume can be easily evaluated:
%--------------------------------------------------------------------
\begin{eqnarray}
\frac{E_{shell}}{L^3}&=&\frac{E - E_{Weyl} + \Delta E}{L^{3}}, \\
\Delta E &=& -\mu  ( \rho_{0} -
                           \rho_{Weyl} )L^3
\nonumber \end{eqnarray}
%--------------------------------------------------------------------
where
the exact and the Weyl (smooth) energy \cite{brack,hilf} per unit
volume are given by
%--------------------------------------------------------------------
\begin{eqnarray}
 \frac{E}{L^3}&=&\frac{2}{L^3}\frac{\hbar^{2}}{2 m a^{2}}
\frac{\pi^{3} }{2}\left ( \frac{L}{a}\right ) ^{2}
\Bigg{[}\frac{1}{4} \left ( \frac{k_{F} a}{\pi} \right
 )^{4} N \\ \nonumber
&- &\frac{N(N+1)(2N+1)(3N^{2}+3N-1)}{120} \Bigg{]}, \\
\frac{E_{Weyl}}{L^{3}}&= &\frac{2}{L^3}
\frac{\hbar^{2}}{2 m a^{2}}
\frac{\pi^{3} }{2}\left ( \frac{L}{a}\right ) ^{2}
\Bigg{[}\frac{1}{5} \left ( \frac{k_{F} a}{\pi} \right ) ^{5}
\nonumber \\
&-&\frac{1}{8} \left ( \frac{k_{F} a}{\pi} \right ) ^{4}  \Bigg{]}
\end{eqnarray}
%--------------------------------------------------------------------
In the above formula
%--------------------------------------------------------------------
\begin{equation}
N={\mathrm {Int}}\left [ \frac{ k_{F} a}{\pi}   \right  ]
\end{equation}
%--------------------------------------------------------------------
stands for the integer part of the argument in the square brackets,
and $a=L - 2 R$ is the distance between slabs and $R$ is the half of
the width of the slab.  Here $L^3$ is the volume of an elementary
(cubic) cell and the factor $'2'$ in front stands for the two spin
states. The average matter density (the number of neutrons per unit
volume) $\rho _0$ and the smoothed density $\rho_{Weyl}$ are
determined by relations
%--------------------------------------------------------------------
\begin{eqnarray}
& &\rho_0= 2\sum _{n=1}^\infty \int \frac{d^2k}{(2\pi )^2}\Theta
\left ( \mu -\frac{\hbar^2k^2}{2m}-
\frac{\hbar ^2n^2\pi^2}{2ma^2}\right ) \\
& &=  \frac{2}{L^3}\frac{\pi}{4}
\left ( \frac{L}{a} \right )^{2}
\left [ \left ( \frac{k_{F} a}{\pi} \right )^{2} N
-\frac{N(N+1)(2N+1)}{6} \right ] . \nonumber \\
& &\rho_{Weyl} = \frac{2}{L^{3}}\frac{\pi}{2}\left ( \frac{L}{a} \right )^2
\Bigg{[} \frac{1}{3}\left ( \frac{k_{F}a}{\pi} \right )^3 -
\frac{1}{4}\left ( \frac{k_{F}a}{\pi} \right )^2 \Bigg{]} \nonumber .
\end{eqnarray}
%--------------------------------------------------------------------
Using these formulas one can show that the shell correction
energy has the behavior
%----------------------------------------------------------------------
\beq
\frac{E_{shell}}{L^3}= \frac{\hbar^2k_{F}^{2}}{40 a^2L m}
G\left  ( \frac{k_Fa}{\pi} \right ),
\eeq
%----------------------------------------------------------------------
where $G(x)$ is an approximate periodic function of its argument, for
$x\ge 1$), $G(x+1)\approx G(x)$, with properties $G(x=n/2)\approx 0$
and approximately $-1\le G(x)\le 1$.  Furthermore
%----------------------------------------------------------------------
\begin{equation}
\rho _{out} = \frac{\rho_0}{v}
\end{equation}
%----------------------------------------------------------------------
is the actual density of the neutron gas between the two slabs and
%----------------------------------------------------------------------
\begin{equation}
v=1-u=\frac{L-2R}{L}
\end{equation}
%----------------------------------------------------------------------

\noindent is the filling factor, which is the ratio of the occupied
volume to the volume of the cell. One can show also that
%----------------------------------------------------------------------
\begin{equation}
\rho _0 = \rho _{Weyl} + \frac{k_F}{12La}F\left
  (\frac{k_fa}{\pi}\right) ,
\end{equation}
%----------------------------------------------------------------------
where $\rho _{Weyl}$ is the Weyl approximation to the
density and $F(x+1)\approx F(x)$ is an approximate periodic function
of its argument too, for $x\ge 1$, with properties $-1\le F(x)\le 0.5$
and $F(x=n)=-1$.  This periodicity leads to the clear pattern of
``valleys'' ($k_Fa =(n+3/4)\pi$) and ``ridges'' ($k_Fa \approx
(n+1/4)\pi$) in the profile of the shell energy shown in
Fig. 1a. These features of the energy and density are naturally
related to fact that these quantities are almost periodic functions in
the classical action along the only periodic orbit in the system,
i.e. in the variable $S=2k_Fa$.

In the case of rod--like and spherical voids we shall use the
semiclassical theory in order to compute the shell energy.  Since we
are interested only in the ``gross shell structure'' we have to take
into account a few of the shortest periodic orbits among the nearest
neighbors only. The lengths of the shortest periodic orbits depend on
the lattice type.  In the following we will assume the simple cubic
and simple square lattices for spherical and rod--like phases
respectively.  The expression for the shell energy density and the
neutron density reads:
%--------------------------------------------------------------------
\begin{eqnarray}
\frac{E_{shell}}{L^{3}} &=&
\frac{1}{L^{3}}
\int_{0}^{\mu} (\varepsilon - \mu) \sum_{i}
g_{shell}(\varepsilon, L_{i})d\varepsilon \\
\rho_0 &=&
\frac{1}{L^{3}}
\int_{0}^{\mu}\left [ g_{Weyl}(\varepsilon ) +
\sum_{i} g_{shell}(\varepsilon, L_{i})\right ] d\varepsilon ,
\end{eqnarray}
%--------------------------------------------------------------------
where $g_{shell}(\varepsilon , L_i )$ denotes the contribution to the
level density due to the orbit $L_i $ and $g_{Weyl}$ is the smooth
level density determined using the Weyl prescription \cite{hilf}.

For the rod--like phase we took into account four orbits of the length
$2L_{1}=2(L-2R)$ and four orbits of the length
$2L_{2}=2(L\sqrt{2}-2R)$. Introducing longer orbits did not lead to
noticeable changes in the patterns presented here. Hence the shell
energy per volume is equal to:
%--------------------------------------------------------------------
\begin{equation}
\frac{E_{shell}}{L^3}=\frac{1}{L^{3}}\int_{0}^{\mu} (\varepsilon- \mu)
\sum _{i=1}^2 A_ig_{shell}(\varepsilon ,L_i) d\varepsilon ,
\end{equation}
%--------------------------------------------------------------------
where $A_1=A_2=4$ and the chemical potential $\mu$ is determined by
the condition:
%--------------------------------------------------------------------
\begin{equation}
\rho_0=\rho _{Weyl}+ \frac{1}{L^3}\int_{0}^{\mu}
\sum _{i=1}^2A_i g_{shell}(\varepsilon ,L_i)
d\varepsilon.
\end{equation}
%--------------------------------------------------------------------
A periodic orbit of the type considered by us gives actually a
contribution with a factor 1/2, since only half of it belongs to a
particular elementary cell. Because there are two spin states, and
thus eight orbits in total, each type of orbit eventually is weighted
by four. The density of states was evaluated using the convolution of
the exact 1--dimensional density of states and the density of states
given by Gutzwiller trace formula for the 2--dimensional system of
disks, which is the cross section of the rod--like system we are
interested in. In some cases such a procedure can lead to spurious
contributions, which are however rather easy to single out, see
Refs. \cite{niall}.  For a given periodic orbit of length $2L_i$, the
shell correction to the density of states is given by the following
expression:
%--------------------------------------------------------------------
\begin{eqnarray} \label{corg}
& & g_{shell}(\varepsilon ,L_i)=\frac{m L L_{i}}{2\pi \hbar^{2} }
\sum_{n=1}^\infty
\frac{\mbox{J}_0(2nkL_i)}{\sinh{n\kappa_i}},  
\end{eqnarray}
%--------------------------------------------------------------------
where the summation is over repetitions of this orbit and
%--------------------------------------------------------------------
\begin{equation}
\varepsilon = \frac{\hbar ^2 k^2}{2 m}.
\end{equation}
%--------------------------------------------------------------------
When one is interested in the gross shell structure then the
contribution of long orbits as well as the contributions due to
repetitions of short primitive orbits vanish under energy averaging.

The explicit form of the shell energy and of the fluctuating part
of the density reads:
%--------------------------------------------------------------------
\begin{eqnarray} \label{core}
& &\frac{E_{shell}}{L^3}=
-\frac{1}{L^3}\frac{\hbar^2k_F^2}{2m\pi}\frac{1}{4}
\sum_{i=1}^2A_i \frac{L}{L_i} \sum_{n=1}^\infty
\frac{{\mbox{J}}_2(2nk_FL_i)}{n^2\sinh(n\kappa_i)}, \\
& &\rho_0=\rho_{Weyl}
          +\frac{k_F}{4\pi L^2} \sum_{i=1}^2 A_i
\sum_{n=1}^{\infty}
\frac{{\mbox{J}}_1(2nk_FL_i)}{n\sinh(n\kappa_i)}.
\end{eqnarray}
%--------------------------------------------------------------------
The parameter $\kappa_{i}$ determines the stability of the orbit
$L_{i}$:
%--------------------------------------------------------------------
\begin{equation}
\kappa_i =
\ln \left [ 1 +\frac{L_i }{R} +
\sqrt{ \frac{L_i }{R}\left ( \frac{L_i }{R}+2 \right ) } \right ] .
\end{equation}
%--------------------------------------------------------------------

\noindent The shell energy as a function of the anti--filling factor
(relative void volume) $u=\displaystyle{\frac{\pi R^2}{L^2}}$ and
$\rho_0$ is shown in Fig 1b. The shell energy has a smaller
amplitude then in the case of the plate--like phase. This comes about
because the periodic orbits are now hyperbolic in the plane
perpendicular to the rods.  Note however that the pattern of
``valleys'' and ``ridges'' looks very similar to the one for the
slabs.  This is because the main contribution due to the classical
orbit of length $2L_{1}$ is the same. There are small interference
effects caused by the orbit of length $2L_{2}$ however. Since it is
longer, this second trajectory contributes with a smaller weight.

For the case of spherical voids there are $26$ periodic orbits between
nearest neighbors of three different lengths $2L_{1}=2(L-2R)$,
$2L_{2}=2(L\sqrt{2}-R)$ and $2L_{3}=2(L\sqrt{3}-R)$.  Thus the shell
energy and density are equal to:
%--------------------------------------------------------------------
\begin{eqnarray}
\frac{E_{shell}}{L^3}&=&\frac{1}{L^{3}}\int_{0}^{\mu}(\varepsilon- \mu )
\sum _{i=1}^3  A_i g_{shell}(\varepsilon ,L_i)
d\varepsilon , \\
\rho_0&=&\rho_{Weyl} + \frac{1}{L^3}\int_{0}^{\mu}
\sum _{i=1}^3 A_i g_{shell}(\varepsilon ,L_i)
d\varepsilon .
\end{eqnarray}
%--------------------------------------------------------------------

\noindent The contribution due to one periodic orbit to the
fluctuating part of the level density reads:
%--------------------------------------------------------------------
\begin{equation}
g_{shell}(\varepsilon ,L_i )=\frac{m L_{i}}{2\pi\hbar^2 k}\sum_{n=1}^{\infty}
\frac{\cos(2 n k L_{i})}{\sinh^{2}(n\kappa_{i})}.
\end{equation}
%--------------------------------------------------------------------
Hence we get:
%--------------------------------------------------------------------
\begin{eqnarray}
& &\frac{E_{shell}}{L^{3}}=\frac{1}{L^{3}}
\frac{\hbar^{2}k_F^2}{2m}\sum_{i=1}^{3}
\frac{A_i}{8\pi (k_F L_{i})^2} \times  \\
& &\sum_{n=1}^{\infty}
\frac{ [ 2 n k_{F} L_{i}
 \cos (2 n k_{F} L_{i}) -
   \sin (2 n k_{F} L_{i}) ]}
{n^{3} \sinh ^{2}(n\kappa_{i} )}, \nonumber \\
& &\rho_0=\rho_{Weyl}+
\frac{1}{L^3}\frac{1}{4\pi}
\sum_{i=1}^{3} A_{i}
\sum_{n=1}^{\infty}\frac{\sin(2 n k_{F} L_{i} )}
{n \sinh ^{2}(n\kappa_{i} )},
\end{eqnarray}
%--------------------------------------------------------------------
where $A_{1}=6, A_{2}=12, A_{3}=8$ respectively.  The shell energy for
the spherical phase is shown in Fig 1c.  In this case the
anti--filling factor is given by $u=
\displaystyle{\frac{4}{3}\frac{\pi R^3}{L^{3}}}$.  A stronger
interference pattern due to the orbits $L_2$ and $L_3$ can be
seen. The amplitude of the shell effects is also lower due to the
greater instability of the orbit on one hand, and due to the smaller
relative volume of the scatterers on the other hand.

In a similar manner one can obtain the interaction energy between two
isolated bubbles at large separations ($a=L-2R\gg R$)
%----------------------------------------------------------------------
\beq
E_{\circ \circ} \approx \frac{\hbar ^2 k_{F} R^2}{8 \pi m}
\frac{\cos (2k_F a)}{a ^3} .
\eeq
%----------------------------------------------------------------------
When compared with the interaction (\ref{r-k}) one observes a similar
behaviour, even though now the two ``impurities'' are large $k_FR\gg
1$. It can be shown however that if one computes instead the same
energy for fixed chemical potential, instead of particle number as was
done here, the bubble--bubble interaction will decay inversely
proportional to the square of the separation \cite{andreas}. 
In a recent paper
\cite{spruch} the Casimir energy for similar arrangements has been
calculated using the semiclassical approximation. In the case of
Casimir energy the situation is somewhat simple, since instead of two
independent dimensionless parameters, $k_FR$ and $k_FL$, only one
dimensionless parameter exists, $R/L$. Thus the Casimir energy for two
spheres has naturally the form
%----------------------------------------------------------------------
\beq
E^{Cas}_{\circ \circ}= \frac{\hbar c}{L} F\left ( \frac{R}{L}\right ),
\eeq
%----------------------------------------------------------------------
with an unknown function $F(x)$. A similar, but much stronger result
can be obtained for the critical Casimir energy \cite{cas}, where one
can show that the theory is conformal invariant.  The authors of
Refs. \cite{spruch} provide also a very compelling argument why the
semiclassical approximation should be particularly accurate for the
calculation of the Casimir energy in case of ideal metallic boundaries
and they show that using only the single periodic orbit the Casimir
energy for two spheres is given by
%----------------------------------------------------------------------
\beq
E^{Cas}_{\circ \circ}= -\frac{\pi ^3 \hbar cR}{720L^2} ,
\eeq
%----------------------------------------------------------------------
and dismiss this result as being valid for large separations, since it
contradicts their expectations that it should agree with the
Casimir--Polder interaction \cite{CP}
%----------------------------------------------------------------------
\beq
E^{CP}_{\circ \circ}\propto -\frac{\hbar c R^6}{L^7} .
\eeq
%----------------------------------------------------------------------
The authors of Ref. \cite{spruch} argue that the contributions arising
from the diffractive paths discussed in Refs. \cite{wirzba}, should
eventually lead to additional contributions, which will cancel exactly
this longer range interaction and in the end, the authors hope that
the Casimir--Polder result will be retrieved. The difference between
these two results for the Casimir energy is very similar to the
difference between the interaction (\ref{r-k}) between two point--like
impurities ($k_FR\ll 1$) and the interaction between two ``fat''
bubbles ($k_FR\gg 1$) \cite{andreas}. In the case of
``fat'' bubbles, the contribution of diffractive orbits are
exponentially small ($\propto \exp (-\alpha k_FR )$, where $\alpha$ is
of order unity) \cite{wirzba}, as one would naturally expect in the
case when rays are a very good approximation to the wave
phenomena. The resolution of this apparent conundrum lies in resolving
the clear clash of limits. When the size of the scatterer $R$
decreases the contribution of the diffractive paths (creeping orbits)
increases and an increasingly larger number of them contribute
significantly to the scattering and thus to the propagator. In the
limit $k_F R\rightarrow 0$ the standard geometric orbit approach has
to be modified, see Ref. \cite{small} and our discussion around
Eqs. (7--10).  It is notable that in the case of the critical Casimir
effect, even longer range interactions ($\propto 1/a^{1+\epsilon}$,
with very small $\epsilon$) between two spheres are possible
\cite{cas}.

The structure of the shell energies shown in the Fig. 1 indicates the
existence of the optimal void sizes (with respect to the shell
effects) for a given outside nucleon density.  Note that for all
phases and for $\rho_0>0.05 fm^{-3}$ the shell energy exhibits a
remarkable softness toward adding additional neutrons to the system
(the ``valleys'' and ``ridges'' are almost horizontal in the Fig. 1).
Hence one can conclude that once the size of the voids have been
determined by minimization of the total energy of the system, an
increase in the number of neutrons outside the voids will not affect
much the shell energy of the system. However, the surface energy will
be affected.

In the Fig. 2 we show the shell energies as a function of $\rho_0$
for the optimal filling factors and nuclear radii determined in
Ref. \cite{oya1}.  One can see that the amplitudes of the shell
energies in the region $\rho_0 \approx 0.04-0.07 $ are of the order of
$10 keV/fm^{3}$, $3 keV/fm^{3}$ and $0.05 keV/fm^{3}$ for
plate--like, rod--like and bubble--like phases, respectively.  There
are usually one or two shallow shell energy minima for the density
range $\rho_0>0.03 fm^{-3}$. The minima are more pronounced in the
case of spherical bubble--like phase mainly due to the stronger
interference effects caused by longer orbits.

Once a phase is formed there is a positional order maintained by the
Coulomb repulsion between spherical nuclei, rods or slabs
\cite{rav,hash,oya,wil,las,lor}.  Although the Coulomb energy is a
smooth function of the void displacement \cite{pet1}, the shell energy
is not.  Since several different orbits contribute to the shell
effects (except for slab--like phase) the displacement of a single
bubble--like or rod--like void from its equilibrium position in the
lattice will give rise to the interference effects.  The interference
pattern will depend on the type lattice. For the simple cubic and
simple square lattices for spherical nuclei and rods, respectively, we
show in the Fig. 3 the changes in the energy due to such ``defects''.
For the plate--like system there is only one direction of displacement
(we do not consider the shear mode) denoted by $x$ perpendicular to
the slab (Fig. 3a).  Since the rod--like phase is a two-dimensional
system, in the Fig 3b we have shown the shell energy as a function of
two perpendicular displacements $x$ and $y$.  They are perpendicular
to the rods and point in the direction of the nearest neighbor.  The
same axes have been chosen for the spherical system although it will
not exhaust all possible directions in the system.  The behavior of
the shell energy in this case is shown in Fig 3c.

The structure of the shell energy surface as a function of a
displacement depends on the lengths of the shortest periodic orbits.
Except for the trivial plate--like phase, in the rod--like and
spherical phase there exist directions into which is easier to locally
deform the lattice.

In Fig. 4 we show the pattern of the energy changes induced by
deforming the rod--like lattice. We considered only volume conserving
deformations. The square lattice was stretched by a factor $\alpha $
in the $x$--direction, by a factor $\beta $ in the $y$--direction and
also the angle between the two axes has been changed to $\gamma$. In
order to preserve the volume all these three parameters should satisfy
the condition
%--------------------------------------------------------------------
\beq
\alpha \beta \sin \gamma =1.
\eeq
%--------------------------------------------------------------------
The case $\alpha =\beta$ and 
$\gamma = \pi /3$ correspond to a perfect triangular lattice.

Increasing the temperature will weaken the shell effects.  At
sufficiently high temperatures the nuclear lattice will disappear. At
smaller temperatures however, when the lattice can be regarded as
frozen, the rise of the temperature will affect mainly shell effects
in the neutron gas. In order to wash out completely the shell effects
the temperature $T$ should be of the order of half of the
distance between shells The spacing between two consecutive shells is
determined by the length of the shortest orbit, $a = L-2R$.  Thus the
energy distance between shells can be determined from the requirement:
$2 k a = 2\pi$ and is given by the expression:
%--------------------------------------------------------------------
\begin{equation}
\Delta E = \frac{\hbar^{2}\pi^2}{2m (L-2R)^2}.
\end{equation}
%--------------------------------------------------------------------
For the optimal filling factors and lattice constants
of various phases obtained in Ref. \cite{oya1}
one obtains the following estimates for the critical temperature:
%--------------------------------------------------------------------
\begin{eqnarray}
T_{c} &\approx& 32 MeV \mbox{ for plate--like system}, \nonumber \\
T_{c} &\approx& 19 MeV \mbox{ for rod--like system},   \\
T_{c} &\approx& 12 MeV \mbox{ for spherical system}.   \nonumber
\end{eqnarray}
%--------------------------------------------------------------------

An accurate description of the shell effects as a function of the
temperature can be obtained using the temperature averaged level
density \cite{brack,kol}:
%--------------------------------------------------------------------
\begin{equation}
g_{shell}(\varepsilon , T)=
\sum_{p.o.}\frac{A_i\tau_{i,n}(T)}{\sinh\tau_{i,n}(T)}
g_{shell}(\varepsilon ,L_i),
\end{equation}
%--------------------------------------------------------------------
where the sum is taken over all periodic orbits including the number
of repetitions $n$ of the orbit and
%--------------------------------------------------------------------
\begin{equation}
\tau_{i,n}=\displaystyle{\frac{2 \pi T m n L_{i}}{\hbar^2 k}}.
\end{equation}
%--------------------------------------------------------------------

Consequently the oscillating part of the free energy density is given by the
formula:
%--------------------------------------------------------------------
\begin{equation}
\frac{F_{shell}}{L^3} = \frac{1}{L^{3}}\int_{0}^{\mu}(\varepsilon - \mu
)g_{shell}(\varepsilon ,T)
 d\varepsilon .
\end{equation}
%--------------------------------------------------------------------
The estimates for different phases are shown in Fig. 5. For simplicity
we have retained in these calculations the value of the Fermi momentum
$k_F$ equal to its zero temperature limit, and therefore the neutron
matter density is not temperature independent in these figures. One
can see that thermal effects will wash out the shell correction energy
at temperatures of the order of 10 MeV and higher.

Now at the end of this analysis we suspect that there are a lot of
other effects, which might be relevant. We did not consider periodic
orbits bouncing between three or more objects. An orbit bouncing
between two bodies leads to a pairwise interaction. Orbits bouncing
between three or more bodies would lead to genuine many body
interactions.  We have also considered only perfectly smooth
objects. If one allows for some degree of corrugation of these
surfaces, many more periodic orbits are likely to appear and that
would lead to even more complicated interactions and more complicated
interference patterns. The fact that corrugation can influence in a
significant, perhaps major way, the Casimir energy, has already been
predicted and measured experimentally \cite{corrugation}. The long
range character of the interaction together with its oscillatory
nature could very easily be at the origin of disorder, even at zero
temperature. At finite temperature disorder is more likely to occur,
due to entropic effects \cite{bub1,bub2}.  We did not consider here
the role of pairing, which we expect however to lead to a certain
flattening of the shell effects \cite{bub1,bub2}, which, however,
should not be interpreted as disappearance of shell effects.
Especially at subnuclear densities neutron pairing should be rather
strong \cite{pairing}. A completely different type of softness of
these structures has been argued in Ref. \cite{pet1}, according to
which the mantles of neutron stars resembles more liquid crystals than
solids.

In the paper we have studied the shell effects in the neutron medium
filled by different nuclear phases.  To our knowledge this is the
first approach which considers specifically the shell effects in the
outside neutron gas and we aimed at discussing its basic features.
Even though in principle Hartree--Fock calculations include in
principle such effects already, the calculations performed so far
\cite{negele} were too narrow in scope and did not address this issue
specifically.  Using semiclassical methods, we have analyzed the
structure of the shell energy as a function of the density, filling
factor, lattice distortions and temperature.  We expect that our
result overestimate somewhat the amplitude of the shell
effects. However, the emerging qualitative overall picture should
remain valid and further microscopic studies are highly desirable. The
main lesson one should remember from this work is that the amplitude
of the shell energy effects is comparable with the energy differences
between various phases determined in simpler liquid drop type models.
The magnitude of the quantum corrections to the ground state energy of
the inhomogeneous neutron matter we have found is significantly larger
than that determined in Ref.\cite{oya2}. The analysis of
Ref.\cite{oya2} was limited however to the motion of nucleons inside
nuclei embedded in a lower density neutron gas.

Our results suggest that the inhomogeneous phase has perhaps an
extremely complicated structure, maybe even completely disordered,
with several types of shapes present at the same time.

The DOE financial support is gratefully acknowledged.  This research
was supported in part by the Polish Committee
 for Scientific Research (KBN) under Contract No. 2 P03B 040 14.  AB
thanks N.D. Whelan, O. Agam, T. Guhr and S.C. Creagh for discussions
and correspondence concerning various aspects of the semiclassical
approximation. PM thanks the Nuclear Theory Group for hosting his
visit in Seattle. AB thanks the members of the Nuclear Theory Group at
the Institute of Theoretical Physics in Warsaw for their
hospitality. And last but not least, AB thanks W.A. Weidenm\"uller for
being such a gracious host.

\end{document}